\documentclass[12pt,a4paper]{article}
\usepackage{a4wide}
\usepackage{latexsym}
\usepackage{epsf}
\usepackage{amssymb}
\def\pd{\partial}
\def\a{\alpha}
\def\b{\beta}

\def\d{\delta}
\def\m{\mu}
\def\n{\nu}
 
\def\l{\lambda}

\def\s{\sigma}
\def\e{\epsilon}

\def\be{\begin{equation}}
\def\ee{\end{equation}}
\def\bea{\begin{eqnarray}}
\def\eea{\end{eqnarray}}

\newtheorem{theorem}{Theorem}

\begin{document}
\begin{flushright}
IFT-UAM/CSIC-01-35\\
hep-th/0111169\\
\end{flushright}

\vspace{1cm}

\begin{center}

{\bf\Large  Dyons, K-theory and M-theory}

\vspace{.7cm}

{\bf C\'esar G\'omez
\footnote{E-mail: {\tt cesar.gomez@uam.es}} and Juan Jos\'e Manjar\'{\i}n}
\footnote{E-mail: {\tt juanjose.manjarin@uam.es}} \\

\vspace{1cm}

{\it   
 Instituto de F\'{\i}sica Te\'orica, C-XVI,
  Universidad Aut\'onoma de Madrid \\
  E-28049-Madrid, Spain}\footnote{Unidad de Investigaci\'on Asociada
  al Centro de F\'{\i}sica Miguel Catal\'an (C.S.I.C.)}

\vskip 1.8cm


{\bf Abstract}
\end{center}
We study the dyonic charge of type IIA D6-branes interpreted as M-theory Kaluza-Klein monopoles. The connection between the dyonic properties of type IIA D6-branes and the existence of eleven dimensions is discussed. We also study dyonic properties of non-commutative monopoles. Variations of the external antisymmetric B-field generate the electric charge of non-commutative dyons.

\newpage
\section{Introduction}

Although the K-theory interpretation of D-brane charges and RR-fields \cite{mm,witten2,mw,hora} provides a solid mathematical ground to Sen's conjecture on open tachyon condensation \cite{sen2}, it also creates new and fascinating problems, as for instance, the deep physical meaning of the underlying gauge theory used to formulate the K-theory interpretation of D-branes.

Mainly motivated by checking string dualities, we have now a very rich understanding of the different uses of D-branes. The interpretation of D-branes as gauge theory solitons, on which the K-theory approach is based, provides the possibility to compare different ways to see the same type of objects. On the other hand, extra dimensions are recently appearing in many different contexts, mainly in the M-theory hypothesis and in the holographic description of gauge theories.

In a previous work, \cite{ces}, it was suggested that the study of dyonic properties of D-branes in type IIA, is a possible way to unravel the dynamical origin of the M-theory extra dimension. In this paper we work out some dyonic properties of monopoles and D-branes in order to seee how far we can go with such a picture.

In section \ref{s2} we consider different aspects of the dyonic properties of KK-monopoles \cite{sen,ghm}, mainly focusing on D6-branes as M-theory KK-monopoles. Our goal is to get some extra insight on the physical meaning of the $E_8$ vector bundle used in the M-theory derivation of K-theory \cite{dmw1}. We also consider NS5-branes from a K-theoretic point of view.

In section \ref{s3}, which can be read independently of section \ref{s2}, we address the problem of dyonic properties of non-commutative monopoles. The main reason for this analysis is the striking similarities between the role played by the B-field in this case and in the example of string KK-monopoles.

\section{Dyons in string theory}\label{s2}

\subsection{KK-monopoles}

As a consequence of S-duality, it is well know \cite{sen,ghm} that Kaluza-Klein (KK) monopoles in string theory become dyons carrying electric winding charge. The physics underlying this phenomena is as follows \cite{ghm,gr}. In ten dimensions a KK-monopole metric is given by ${\mathbb{R}}^{5,1}\times{\cal{N}}_4$, where ${\cal{N}}_4$ is the self-dual euclidean  Taub-Nut space. The boundary of ${\cal{N}}_4$ at spatial infinity is the 3-sphere, $S^3$, interpreted as the Hopf fibration of $S^1$ on $S^2$. In string theory we can consider winding states on $S^1$ which, due to the fact that $\pi_1(S^3)=0$, can unwind in the presence of a KK-monopole. Conservation of winding charge implies that the KK-monopole must carry winding charge\footnote{For a K-theory discussion of this phenomena, see reference \cite{mms}}. After standard KK reduction this winding charge becomes electric charge for the $B_{\m 4}$ $(\m=1,2,3)$ component of the B-field along the compact $S^1$ direction.

A more direct way to understand the dyonic charge is by counting bosonic zero modes of the KK-monopole solution. In addition to the 3 bosonic zero modes corresponding to the center of mass location in ${\mathbb{R}}^3$ we have an extra zero mode associated with pure gauge transformations of the B-field that are non-vanishing at infinity.

For the KK-monopole solution \cite{sor,gp}

\be
ds^2=dy^2+ds^2_{TN}
\ee

\noindent with $ds^2_{TN}$ the self-dual Taub-Nut metric \cite{gh}

\be
\label{taubnut}
ds^2_{TN}=U\left[ dx_4+4m\left(1-\cos\theta\right)d\phi\right]^2+U^{-1}\left[dr^2+r^2\left(d\theta^2+\sin^2\theta d\phi^2\right)\right]
\ee

\noindent where $U=\left( 1+\frac{4m}{r}\right)^{-1}$, we can define a pure gauge B-field \cite{sen}

\be
\label{pgf}
B=\a d\left(\frac{r}{r+4m}\s_3\right)
\ee

\noindent with $\s_3=\frac{1}{4m}\left[dx_4+4m(1-\cos\theta)d\phi\right]$. Obviously, the pure gauge field (\ref{pgf}) is non vanishing at $r\rightarrow\infty$. Moreover, it is a zero mode since it is proportional to the unique harmonic two form in euclidean Taub-Nut space.

As already pointed out in references \cite{sen,ghm}, the previous derivation of the dyonic charge of KK-monopoles in string theory is very much similar to the characterization of the moduli space of BPS-monopoles in $SU(2)$ Yang-Mills-Higgs gauge theories. In this case, the pure $U(1)$ gauge transformation $e^{i\a\Phi(\vec x)/|\Phi(\vec x)|}$, with $\Phi$ the Higgs field and $\a\in[0,2\pi]$ is non vanishing at infinity in the monopole background. The parameter $\a$ defines the coordinate on the $S^1$ fiber of the moduli space ${\cal{M}}_1={\mathbb{R}}^3\times S^1$ and the motion along this fiber generates the electric charge of the monopole that is defined by the corresponding conjugate momentum.

In the KK case we observe that continous changes of $\a$ in (\ref{pgf}), i.e. a one parameter family of type $B(t)=\a_0td\left(\frac{r}{r+4m}\s_3\right)$ generates a non vanishing $H$ field with winding charge proportional to the conjugate momentum $\a_0$ \cite{sen}.

The analogy between the dyon effect for the KK-monopoles in string theory and the dyonic charge for BPS-monopoles can be pushed a bit further. For instance, in the BPS case for the 't Hooft-Polyakov $SU(2)$ monopole, the relevant homotopy groups are

\be
\pi_3\left(SU(2)\right)=\pi_2\left(SU(2)/U(1)\right)=\pi_1\left(U(1)\right)={\mathbb{Z}}
\ee

The gauge transformation $e^{i\a\Phi(\vec x)/|\Phi(\vec x)|}$ we have used to generate the dyon zero mode is for $\a=2\pi$ a non trivial gauge transformation in $\pi_3\left(SU(2)\right)$ with winding number equal to minus the magnetic charge of the monopole. This fact is crucial in order to derive Witten's relation between the dyon electric charge and the instanton $\Theta$ vacuum angle \cite{witten1}.

The question that now arises is what is playing the analogous role of $\pi_3$ in the KK-monopole?. In order to address this question we will first consider the case of KK-monopoles in the heterotic string. The analogous role to $\pi_3$ should correspond to the winding

\be
\label{wn}
\frac{1}{16\pi^2}\int_{S^3}\omega
\ee

\noindent for $S^3$ the boundary defined by the Hopf fibration and $\omega$ the Lorentz-Chern-Simons 3-form\footnote{This quantity is simply the NUT-charge of the gravitational instanton}. Since we are working in the heterotic string, the Lorentz-Chern-Simons form $\omega$ contributes, by anomaly cancellation arguments \cite{gs}, to the field strength of the antisymmetric tensor field $B_{\m\n}$ and (\ref{wn}) becomes proportional to the H-charge $\int_{S^3}H$. Moreover, the H-field defined by $\omega$ is equal to $K\wedge B$ for $B$ given by (\ref{pgf}) with $\a=2\pi$ and where $\int_{S^1}K=2\pi R$ is the lenght of the $S^1$ fiber at infinity.

Notice that the winding number (\ref{wn}) only depends on the Taub-Nut geometry and is the same for all string KK-monopoles, however it is only directly related to the field strength of the antisymmetric tensor field and therefore to the B-field gauge transformation used to define the dyon zero mode, in the heterotic case where cancellation of anomalies forces the know relation between $H$ and the Lorentz-Chern-Simons term\footnote{Let us just notice that, in the heterotic case, the KK-monopole carries -1 unit of H-monopole charge \cite{sen}, due to the same relation between $H$ and the Lorentz-Chern-Simons form. This makes topologically the KK-monopole in this case very close to the BPS $SU(2)$ monopole. Namely, the $SU(2)$ vector bundle of a BPS-monopole on the boundary $S^2$ is isomorphic to $H\oplus H^{-1}$ for $H$ the Hopf bundle. We can relate these two Hopf bundles to the KK and H-monopole charges respectively.}

\subsection{D6-branes and KK-monopoles}

The analogy between KK-monopoles and BPS-monopoles is potentially richer in the case of D6-branes. In fact, following Sen's approach \cite{sen2} to D-branes as solitons for open tachyon condensation, we can consider the D6-brane as a real BPS-monopole of the 't Hooft-Polyakov type, for a ten dimensional $U(2)$ gauge theory spontaneously broken to $U(1)\times U(1)$ with the open tachyon field playing the role of Higgs field. In addition we can think of the D6-brane as a KK-monopole for M-theory \cite{town}. Which lead in \cite{dl} to a reinterpretation of the four dimensional electric/magnetic duality as a particle/D6-brane duality in ten dimensions.

Let us first briefly review the solitonic description of D6-branes. The relevant non BPS configuration of D-branes is defined in terms of two D9 filling branes in type IIA string theory. The low energy world volume theory is a ten dimensional $U(2)$ gauge theory with the open tachyon transforming in the adjoint representation. Assuming spontaneous breaking $U(2)\rightarrow U(1)\times U(1)$ for a diagonal vev of the open tachyon, the relevant topology defining the D6-branes as a bound state is

\be
\pi_2\left(U(2)/U(1)\times U(1)\right)=\pi_1\left(U(1)\right)={\mathbb{Z}}
\ee

\noindent which parametrizes the stable tachyon vortices. The tachyon field is given by \cite{hora}

\be
\label{taq}
T=\sum_{i=1}^3x^i\s^i
\ee

\noindent with $\s^i$ the Pauli matrices and $x^i$ coordinates on the transversal ${\mathbb{R}}^3$. Notice that $T$ in (\ref{taq}) defines a map from $S^2=\left\{ x^i\in{\mathbb{R}}^3:\sum x_i^2=1\right\}$ into $U(2)/U(1)\times U(1)$ with winding number equal to one.

From the point of view of K-theory \cite{witten2,hora} the D6-brane is related to the relative K-group $K^{-1}\left(B^3,S^2\right)$ which is isomorphic to $\tilde K^{-1}\left( S^3\right)$ and thus\footnote{Remember that the higher K-groups satisfy $K^{-m}\left(X\right)=K\left(\Sigma^mX\right)$, where $\Sigma^mX\equiv S^m\wedge X$ is the reduced suspension of the topological space $X$, which, in the $n$-sphere case, and for $m=1$, is simply $\Sigma S^n=S^1\wedge S^n\sim S^{n+1}$.} to $\tilde K\left( S^4\right)$, see for example \cite{os}. The group $\tilde K\left( S^4\right)$ is characterized by the stable homotopy group $\pi_3\left( U(2)\right)$ that is equal to $\pi_2\left(U(2)/U(1)\times U(1)\right)$.

The dyonic nature of the D6-brane as a 't Hooft-Polyakov monopole for the open tachyon condensation derives from the fact that the gauge transformation $e^{i\a T\left(\vec x\right)}$ is non vanishing at infinity and for $\a=2\pi$ is topologically non trivial in $\pi_3\left(U(2)\right)$. 

It is worth to point out that the relevant topology to the dyonic nature of the monopole is precisely the same appearing in the K-theory description, namely, the $\pi_3\left(U(2)\right)$, which characterizes\footnote{Notice, from a more formal point of view, that $K^{-1}(X,Y)$ is defined in terms of couples $(E,\a)$ where $E$ is a vector bundle on $X$ and $\a$ is an automorphism of $E$ such that $\a/Y$ is homotopic to the identity within the automorphism of $E$. What is interesting to point out is that this automorphism $\a$ that enters into the formal definition of $K^{-1}(X,Y)$ is intimately connected with the definition of the electric charge of the monopole. Namely, $\a$ are rotations around the direction defined by the Higgs open tachyon field. See also reference \cite{hora}.} $\tilde K\left(S^4\right)$.

Let us now consider the D6-brane from a different point of view, namely, as a KK-monopole for M-theory. In this case the extra $S^1$ of the eleven dimensional M-theory space-time is Hopf fibered. 

The previous K-theory description of D6-branes as 't Hooft-Polyakov  $U(2)$ monopoles imply that they can carry dyonic charge. Our goal now will be to understand this dyon charge from the point of view of D6-branes as M-theory KK-monopoles.

Similarly to the case of string KK-monopoles, we should look for some non trivial gauge transformations non vanishing at infinity. In the case of M-theory KK-monopoles, these transformations are going to be gauge transformations of the 3-form field $C$. 

In order to find the pure gauge 3-form $C$ associated with the dyon degree of freedom, we will start with a slightly different version of the Taub-NUT metric

\be
ds^2=\frac{r+m}{r-m}dr^2+(r^2-m^2)d\Omega_{(2)}^2+\left( 4m\right)^2\frac{r-m}{r+m}\left(dx^4+\frac{\cos\theta}{2}d\phi\right)^2,
\ee

\noindent which is related to (\ref{taubnut}) by making the following change of variables

\bea
x^4=\frac{1}{4m}\tilde x^4\pm\frac{1}{2}\phi,\qquad r=\tilde r+m,
\eea

\noindent where the $\pm$ sign stands for the possibility of choosing one of the two coordinate patches. 

We can now construct a zero mode 1-form \cite{yi} as a solution to the equation

\be
\label{zm}
\nabla^\a\nabla_{[\a}A_{\m]}=0.
\ee

\noindent Taking a gauge field of the form

\be
A_{\m}=\left(f_1(r),0,\frac{\cos\theta}{2}f_2(r),f_2(r)\right),
\ee

\noindent it is easy to see that the $f_1(r)$ is not determined by (\ref{zm}) and could be understood as a gauge parameter, however we will determine it below as a result of a Bianchi indentity, and $f_2(r)$ turns out to be 

\be
\label{f2}
f_2(r)=-\frac{1}{\left(2\pi\right)^4l_s^{7/2}}\frac{r-m}{r+m}.
\ee

\noindent As a notational comment, from now on, we will denote the numerical factor in (\ref{f2}) as $k$.

Now we can construct the 3-form Hodge dual to this 1-form following the usual rules of differential geometry, namely

\be
C_3={^*}A=\frac{\sqrt{|g|}}{2}A_\m g^{\m\eta}\epsilon_{\eta\a\b\gamma}dx^\a\wedge dx^\b\wedge dx^\gamma,
\ee

\noindent which, taking explicitely the patch with the minus sign in the original variables, takes the form

\be
\label{3f}
C_3=\s_m\wedge d\theta\wedge d\phi,
\ee

\noindent where

\be
\s_m=\frac{k'}{2}\sin\theta dx^4-\frac{k}{(4m)^2}dr,
\ee

\noindent where, in order to determine the precise form of the function $f_1(r)$, we have imposed the Bianchi identity $dC_3=0$, which leads to the value $f_1(r)=\frac{k'}{r^2}$.

Notice that, effectively, this 3-form corresponds to a non-vanishing pure gauge deformation at infinity, that is, in the $r\rightarrow\infty$ limit, which, in analogy with the previous section, we may write as

\be
\label{a}
C_3=\a\omega_3.
\ee

\noindent with $\a$ a constant parameter and $\omega_3$ given by the previous expression.

In order to make further contact with the previous section, the parameter $\a$ in (\ref{a}) must be a variable in $[0,2\pi]$, which means that we must fix the constants of the three form so that when integrated over a three cycle dual to $\omega_3$, the imaginary part of the action is given by $i\a$.

The constant to be determined is $k'$ which, taking a periodicity of $16\pi m$ in the $x^4$ variable, is equal to 

\be
k'=\frac{9}{(4\pi)^4mR^2l^{3/2}}.
\ee

\noindent where $R$ simply denotes the radius of the $S^3$ where we perform the integral.

We can now see again what happens when we allow a continuous change in $\a$, that is, we allow a parametrization of the form $\a(t)$. Letting $\a(t)=\a_1t+\a_0$, we get a non-vanishing field strength $G_4$ of the form

\be
G_4=dC_3=\frac{\a_1k'}{2}\sin\theta dt\wedge dx^4\wedge d\theta\wedge d\phi-\frac{\a_1k}{(4m)^2}dt\wedge dr\wedge d\theta\wedge d\phi,
\ee

\noindent so the term proportional to $dt\wedge dx^4\wedge d\theta\wedge d\phi$ indicates the existence of a ``{\sl membrane winding charge}'' proportional to $\a_1$. In the context of \cite{sen,ghm} this winding charge was understood as coming from a string winding around the $x^4$ direction, but now it could be understood as coming from a membrane wrapping around the compact extra dimension.

Again this dyonic charge is topologically related to a non vanishing winding number in $\pi_3$. After appropiated normalization, this winding number is simply $\int_{S^3}C$ for $\a=2\pi$. As it was the case for heterotic string KK-monopoles this quantity has a nice gauge theory meaning. In fact, if we assume that the M-theory 3-form $C$ can be written as a Chern-Simons 3-form for a $E_8$ gauge field theory plus the gravitational Chern-Simons term \cite{dmw1,hw,witten3,dmw2} then we can easily relate $\int_{S^3}C$ to the topology of the M-theory $E_8$ vector bundle. In fact, if we work with the strength 4-form $G$ of $C$ and use the relation  \cite{witten3}

\be
\left[\frac{G}{2\pi}\right]=w(V)-\frac{\l}{2}
\ee

\noindent where $\l=p_1/2$, $\left[G/2\pi\right]$ denotes the cohomology class of $G/2\pi$ and $w(V)$ is the second Chern class of the $E_8$ vector bundle V, we observe that for a four dimensional euclidean Taub-Nut space, which has vanishing first Pontryagin number, the cohomology class of $G$ is completely determined by the M-theory $E_8$ vector bundle. Thus we are tempted to conjecture that the $\pi_3$ associated with the dyonic properties of the KK-M-theory monopole are directly related with the second Chern class of the formal $E_8$ bundle defined on the M-theory eleven dimensional space-time.

\subsection{K-theory and 5-branes}

Let us consider the ten dimensional $U(2)$ gauge theory used in the K-theory description of D6-branes. The first thing to be noticed is that, for vanishing vacuum expectation value of the open tachyon field, we have, due to the fact that $\pi_3\left(U(2)\right)={\mathbb{Z}}$, topologically stable extended objects of space codimension four with finite energy density. These objects correspond to twisted gauge configurations which tend to a pure gauge on the $S^3$ sphere at infinity of non vanishing winding number in $\pi_3\left(U(2)\right)$. It is quite natural to identify these objects with the NS5-brane of type IIA.

The charge of the 5-brane in type IIA is given by the integral

\be
\int_{S^3}H
\ee

\noindent of the 3-form $H$. Thus if we identify this NS5-brane with the topologically stable extended object derived from the $\pi_3\left(U(2)\right)={\mathbb{Z}}$ of the K-theory $U(2)$ gauge theory, we will get an identification between $H$ and the gauge Chern-Simons 3-form $w_3(A)$ of the U(2) gauge theory, namely

\be
\label{cc}
\int_{S^3}H=\int_{S^3}w_3(A)=\int_{{\mathbb{R}}^4}Tr\left( F\wedge F\right)
\ee

\noindent where $F\wedge F$ is the second Chern class of the $U(2)$ gauge theory.

From the point of view of the M-theory it is natural to associate the M5-brane with the $\pi_3\left(E_8\right)={\mathbb{Z}}$ for $E_8$ the formal vector bundle defined on the eleven dimensional M-theory space-time manifold. The M-theory version of relation (\ref{cc}) becomes

\be
\label{mcc}
\int_{S^4}G=\int_{S^4}Tr\left(F\wedge F\right)
\ee

\noindent where now $G$ is the strength 4-form of M-theory and $Tr\left( F\wedge F\right)$ is the second Chern class of the M-theory $E_8$ vector bundle.

From (\ref{mcc}) we observe that the non trivial topology of the M5-brane, from the M-theory $E_8$ gauge theory point of view, is associated with non trivial gauge transformations of the 3-form $C$ on the equator $S^3$ of the four sphere $S^4$.

Notice that in both gauge theory descriptions, type IIA or M-theory, of the 5-brane, the crucial step is to identify either $H$ or $C$ with Chern-Simons 3-forms.

\begin{figure}[h]
\begin{center}
\leavevmode
\epsfxsize=12cm
\epsffile{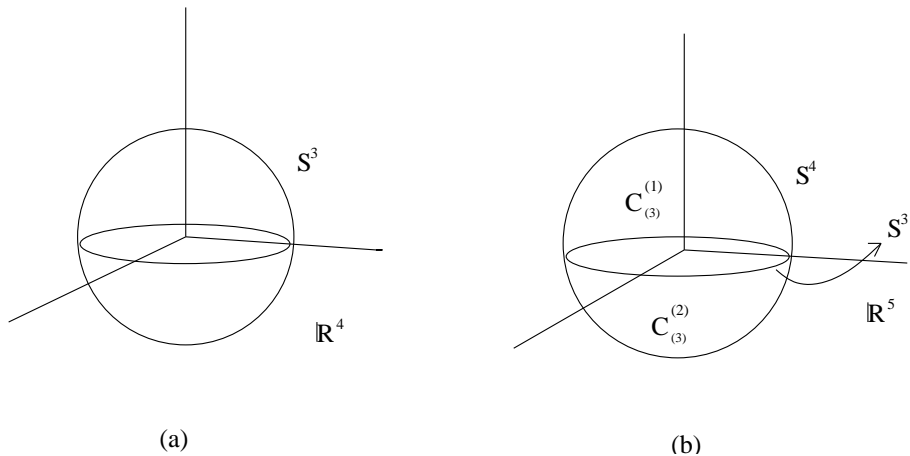}
\caption{\it (a) Configuration of the NS5-brane in type IIA. (b) Configuration of the M5-brane of M-theory.}
\label{f1}
\end{center}
\end{figure}

In figure \ref{f1} we have simbolically represented the topology associated with the NS5-brane of type IIA and the one associated to the M5-brane. In figure \ref{f1}.a, corresponding to the type IIA picture, we have a non trivial $H$ twisted on $S^3$, while in figure \ref{f1}.b, corresponding to the M-theory picture, we have a gauge transformation of the 3-form $C$ on the equator $S^3$.

In reference \cite{ces} it was suggested that the existence of the extra M-theory dimension is intimately connected with the dyonic properties of D-branes in type IIA. The na\"\i ve way to understand this suggestion would be to interpret the D6-branes as $U(2)$ 't Hooft-Polyakov monopoles and directly associating the $S^1$ fiber of the monopole moduli space{\footnote{More generally, the $S^1$ fiber of the multimonopole space ${\mathcal{M}}_k={\mathbb{R}}^3\times\frac{S^1\times {\mathcal{M}}_k^0}{{\mathbb{Z}}_k}$}} ${\mathcal{M}}_1={\mathbb{R}}^3\times S^1$ with the M-theory $S^1$. This is certainly too na\"\i ve since the D6-brane is not really allowed to ``move'' in the extra eleven dimension.

However, this is not the case for the NS5-brane of type IIA, that is the direct dimensional reduction of the M5-brane. The collective RR mode of the NS5-brane in type IIA, that is the ten dimensional signal of the existence of the extra eleven dimension, is associated with gauge transformations of the $C$-field non vanishing at infinity \cite{mms}, precisely these are the gauge transformations of the 3-form $C$-field we have used in equation (\ref{a}) to characterize the dyonic degree of freedom of the D6-brane as a M-theory KK-monopole. Simbolicaly, what is happening is that the dyonic $S^1$ of the D6-brane appears as the extra eleven dimension where the NS5-brane is ``moving''.

The connection between the M-theory extra dimension and the topology of monopole moduli space can be considered from a different point of view. Namely, if we consider $k$ monopoles, the moduli space

\be
\label{mmms}
{\mathcal{M}}_k={\mathbb{R}}^3\times\frac{S^1\times {\mathcal{M}}_k^0}{{\mathbb{Z}}_k}
\ee

\noindent is such that \cite{ah}

\be
\pi_1\left({\mathcal{M}}_k\right)={\mathbb{Z}}
\ee

\noindent which comes from two facts. First, in a particle like approximation we will get $\left({\mathcal{M}}_1\right)^k$, with $\pi_1={\mathbb{Z}}^k$. Secondly, $\pi_1\left({\mathcal{M}}_k^0\right)={\mathbb{Z}}_k$. Thus we get from (\ref{mmms}) $\pi_1\left({\mathcal{M}}_k\right)={\mathbb{Z}}$. For a KK-monopole of charge $k$, the boundary looks like $S^3/{\mathbb{Z}}_k$, with $\pi_1\left( S^3/{\mathbb{Z}}_k\right)={\mathbb{Z}}_k$. On the other hand, for each KK-monopole we have an harmonic 2-form which produces ${\mathbb{Z}}^k$. When we combine the two facts, like in the characterization of $\pi_1\left({\mathcal{M}}_k\right)$, we get ${\mathbb{Z}}$ as a final result (see \cite{mms} for a discussion in terms of singletons).

Let us now consider 5-branes in type IIB. In this case we have a solitonic NS5-brane and a D5-brane, which are related by S-duality. The interesting aspect of this example is that we can consider both types of 5-branes as solitonic. 

From the K-theory point of view, the D5-brane is a soliton for a $U(2)\times U(2)$ gauge theory spontaneously broken to $U(2)_D$ with the open tachyon playing again the role of the Higgs field. The relevant topology for an extended object of space codimension four is

\be
\pi_3\left(U(2)\times U(2)/U(2)_D\right)={\mathbb{Z}}
\ee

The NS5-brane is a soliton with non vanishing charge

\be
\int_{S^3}H_{mnp}=-\int_{S^3}\e_{mnp}^{\hspace{0.6cm}q}\pd_q\Phi
\ee

\noindent for $\Phi$ the dilaton field. On the other hand, the charge of the D5-brane is

\be
\int_{S^3}H^{RR}_{mnp}=\int_{S^3}w(T)
\ee

\noindent where $w(T)$ is the winding number for the open tachyon field

\be
T=\sum_{i=1}^4\Gamma^ix^i
\ee

\noindent and is known to be 1. Since both are S-dual, we can think that the S-dual transformation of $\int_{S^3}\e_{mnp}^{\hspace{0.6cm}q}\pd_q\Phi$ is precisely given in terms of the tachyon field as $\int_{S^3}w(T)$ . The S-dual of the $\Phi$ field is

\be
e^{-\Phi}\rightarrow \frac{e^{-\Phi}}{{C^{(0)}}^2+e^{-2\Phi}}
\ee

\noindent so we can write

\be
\frac{{C^{(0)}}^2-e^{-2\Phi}}{{C^{(0)}}^2+e^{-2\Phi}}\pd_q\Phi+\frac{2C^{(0)}}{{C^{(0)}}^2+e^{-2\Phi}}\pd_qC^{(0)}=\frac{\e^{mnp}_{\hspace{0.6cm}q}}{(24\pi)^2}Tr\left(T^{-1}dT\right)^3
\ee

\noindent which takes a more familiar form in the case $C^{(0)}=0$, so

\be
\pd_q\Phi=-\frac{\e^{mnp}_{\hspace{0.6cm}q}}{(24\pi)^2}Tr\left(T^{-1}dT\right)^3
\ee

This is an interesting relation between the dilaton (closed sector) and the tachyon field (open sector) living on the filling brane. Moreover, since we are working in type IIB, we expect to work with the complexified string coupling constant, which is consistent with the fact that the tachyon field we are using in type IIB is a complex field.

Finally, let us just point out that it is very natural to associate the heterotic 5-brane with the non vanishing $\pi_3\left( E_8\right)$. However, in this case the gauge theory is part of the closed string spectrum of the heterotic string. Probably a heterotic origin of closed string gauge fields can be a good hint to unravel the deep physical meaning of the gauge theories appearing in the K-theory description of D-branes.

\section{Non-commutative dyons}\label{s3}

In this section we will briefly consider the dyon effect for non-commutative monopoles. Our approach will consist in working out the non commutative deformation of the scattering functions used to characterize the magnetic monopole moduli space in the context of Donaldson's theorem. Our result simply shows that continous changes of the external B-field generate motions on the $S^1$ fiber of the monopole moduli space ${\mathcal{M}}_k$, i.e. they generate the dyonic electric charge of the monopole, which can be seen as an Ampere's law. This is reminiscent of the way the dyon charge appears in string KK-monopoles.

The standard construction of BPS monopoles is based on Nahm's extension \cite{nahm} of the ADHM formulation of instantons \cite{adhm}. For $SU(2)$ monopoles of magnatic charge $k$, Nahm's construction can be summarized in the following theorem due to Hitchin \cite{hitchin}

\setcounter{theorem}{0}
\begin{theorem}[Hitchin]\label{h}
BPS-$SU(2)$ monopoles of charge $k$ are one to one related to $O(k;\mathbb{R})$ conjugacy classes of $k\times k$ matrices $T_i(s)$ for $i=1,2,3$ and $s\in [0,2]$ such that

\begin{enumerate}

\item $\label{h1}\frac{dT_i}{ds}+\frac{1}{2}\e_{ijk}[T_j,T_k]=0$.

\item $\label{h2} T_i(s-2)=T_i(s)^T$.

\item $\label{h3} T_i^*(s)=-T_i(s)$.

\item The $T_i(s)$ are smooth functions on (0,2) with simple poles at $s=0,2$.

\item The residues of the $T_i(s)$ at $s=0$ define an irreducible representation of $SU(2)$.

\end{enumerate}

\end{theorem}

The monopole configuration is obtained from the $T_i(s)$ by the ADHM construction for the Dirac operator

\be
\Delta(x)=\left( x_0+\sum x_je_j\right)+i\frac{d}{ds}+i\sum T_je_j
\ee

\noindent the $L^2$ kernel of which can be shown to be two dimensional, so we can obtain the Higgs field and the connection associated to the monopole by choosing an orthonormal basis $(v_1,v_2)$ and performing the following integrals

\be
\Phi(v_a)=\sum_{b=1}^{2}\int_{-1}^{1}(v_b,sv_a)ds
\ee
\be
A_i(v_a)=\sum_{b=1}^{2}\int_{-1}^{1}(v_b,\frac{dv_a}{dx^i})ds
\ee

\noindent where the limits are given by the eigenvalues of the asymptotic Higgs field.

A simple way to derive the equations (\ref{h1}) in theorem (\ref{h}), the so-called Nahm's equations, is as the anti self-duality condition for the field strength $F=dA+A\wedge A$, with {\sl{gauge}} connection 

\be
\label{dc}
A=T_1(s)dp^1+T_2(s)dp^2+T_3(s)dp^3
\ee

\noindent on momentum space ${\mathbb{R}}^4=\lbrace(s,p^1,p^2,p^3)\rbrace$ being the $p_i$'s dual to the original coordinates $x_i$ of ${\mathbb{R}}^3$.

A natural way, from the string theory point of view, to derive a ``non-commutative" deformation of Nahm's equations is to study the anti self-duality for the field ${\cal{F}}=F+\frac{B}{2\pi\a'}$ for $B$ an antisymmetric B-field defined on momentum space. For $B=\theta_{ij}dp^i\wedge dp^j$ with $\theta_{ij}=-\theta_{ji}$ this condition reads

\be
\frac{dT_i}{ds}+\frac{1}{2}\e_{ijk}\left[\left[T^j,T^k\right]+\theta^{jk}\right]=0
\ee

For the particular case $\theta_{12}=\e_{12}\theta$, $\theta_{ij}=0$ for $i,j\neq 1,2$, we get

\be
\label{dene}
\frac{dT_i}{ds}+\frac{1}{2}\e_{ijk}\left[T^j,T^k\right]+\theta\delta_{i3}=0
\ee

\noindent which is the non commutative deformation suggested by Bak in \cite{bak}, while studying the deformation produced by the star product in the integrability conditions of Nahm's construction to first order in $\theta$, and by Gross and Nekrasov in \cite{gn}, who performed a shift in the commutative $T_i(s)$ by means of the matrix coordinates of a noncommutative space-time, namely, ${\cal{T}}_i=T_i+x_i$.

\subsection{The moduli space}

As it is well known, the moduli space ${\cal{M}}_k$ of $SU(2)$ BPS monopoles of magnetic charge $k$ is an hyperk\"ahler complex manifold of dimension $4k$, \cite{ah}, which means that there are three covariantly constant complex structures satisfying the quaternionic algebra. 

The space ${\cal{M}}_k$ is a $S^1$ bundle on the moduli space ${\cal{N}}_k$ of gauge equivalent Yang-Mills-Higgs configurations $\left(A_i(\vec x),\Phi(\vec x)\right)$ satisfying Bogomolny equations

\be
\label{bog}
D\Phi={^*}F
\ee

\noindent on ${\mathbb{R}}^3$. 

The first homotopy group $\pi_1\left({\cal{M}}_k\right)$ is non trivial and equal to $\mathbb{Z}$. Thus for the simplest case $k=1$ we have ${\cal{M}}_1={\mathbb{R}}^3\times S^1$, where points in ${\mathbb{R}}^3$ represent the position of the monopole and the motion in $S^1$ gives rise to the electric charge of the monopole. For generic $k$ the motion of the centre of mass of the system and the total phase decople from the relative motion, so there is an isometric splitting

\be
{\cal{M}}_k={\mathbb{R}}^3\times\frac{S^1\times{\cal{M}}_k^0}{{\mathbb{Z}}_k}
\ee

\noindent with ${\cal{M}}_k^0$ of dimension $4(k-1)$ the reduced monopole space which parametrizes the monopoles with fixed centre and where ${\mathbb{Z}}_k$ acts on the $S^1$ coordinate $\phi$ as $\phi\rightarrow\phi+\frac{2\pi}{k}$.

Given an isomorphism ${\mathbb{R}}^3\sim{\mathbb{C}}\times{\mathbb{R}}$, which is to fix a direction in $\mathbb{R}^3$, there exists a one to one relation between the moduli space ${\cal{M}}_k$ and the space ${\cal{R}}_k$ of based rational functions of degree $k$, $f:\mathbb{CP}^1\rightarrow\mathbb{CP}^1$, and such that $f(\infty)=0$ \cite{don}. This important theorem due to Donaldson associates to each point $m\in{\cal{M}}_k$ a rational function $S_k(z)$ which is the scattering function \cite{ah} for this monopole configuration along the fixed direction.

The procedure to determine $S_k(z)$ for each point $m\in{\cal{M}}_k$ is as follows. Let us first introduce a new matrix $T_0$ and let us impose anti self-duality for $F=dA+A\wedge A$ with $A=T_0ds+\sum_iT_i(s)dp^i$. Then, Nahm's equations become

\be
\label{dne}
\frac{dT_i}{ds}+\frac{1}{2}\e_{ijk}[T^j,T^k]+[T_0,T_i]=0
\ee

Now let us choose the isomorphism ${\mathbb{R}}^3\sim{\mathbb{C}}\times{\mathbb{R}}$, for instance $(x_1,x_2,x_3)\rightarrow (z=x_2+ix_3,x_1)$. Using this isomorphism we define a complex coordinate $w=s+ip^1$. Defining

\bea
\a=\frac{1}{2}\left(T_0+iT_1\right),\qquad \b=\frac{1}{2}\left(T_2+iT_3\right)
\eea

\noindent Nahm's equations (\ref{dne}) become equivalent to

\bea
\label{hne1}\frac{d\b}{ds}+2[\a,\b]=0\\
\label{hne2}\frac{d}{ds}\left(\a+\a^*\right)+2\left([\a,\a^*]+[\b,\b^*]\right)=0
\eea

These equations determine the hyperk\"ahler structure of the moduli space in the sense of \cite{kron}. Namely, the hyperk\"ahler moment maps are defined by

\bea
\m_1(\a,\b)\rightarrow -4i\frac{d}{ds}\left(\a+\a^*\right)\\
\m_2(\a,\b)\rightarrow -4\frac{d}{ds}\left(\b-\b^*\right)\\
\m_3(\a,\b)\rightarrow -4i\frac{d}{ds}\left(\b+\b^*\right)
\eea

\noindent with $\m_2$ and $\m_3$ we can form the following combination $\m_C=\m_3+i\m_2$, which is the complex momentum map, and $\m_R=\m_1$, is the real moment map. Then the moduli space can be identified, via the hyperk\"ahler quotient construction, with

\be
\left(\m_C^{-1}(0)\cap\m_R^{-1}(0)\right)/\mathcal{G}
\ee

\noindent where $\mathcal{G}$ is the Lie group acting on the moduli space preserving the metric and the complex structures.

Equation (\ref{hne1}) is invariant with respect to gauge transformations

\bea
\b\rightarrow g\b g^{-1}\\
\a\rightarrow g\a g^{-1}-\frac{1}{2}\frac{dg}{ds}g^{-1}
\eea

\noindent for $g:(0,2)\rightarrow GL(k,{\mathbb{C}})$ such that $g(2-s)=g^T(s)^{-1}$. 

We define a Nahm complex \cite{don} as $(\a,\b,v)$ with $\a$, $\b$ $k\times k$ matrices satisfying (\ref{hne1}) and

\bea
\nonumber \a(2-s)=\a^T(s),\\
\b(2-s)=\b^T(s),
\eea

\noindent and such that $\a$, $\b$ are smooth on $(0,2)$ with simple poles at $s=0$ and $s=2$ with residues $a$, $b$ at $s=0$ such that $Tr(a)=0$ and

\be
av=\frac{(k-1)}{4}v
\ee

\noindent for $v\in{\mathbb{C}}^k$ and $\left\{ b^iv\right\}_{i=0}^{k-1}$ a basis of ${\mathbb{C}}^k$.

Two Nahm complex $(\a,\b,v)$ and $(\a',\b',v')$ are equivalent if there exist a $g(s)\in GL(k,{\mathbb{C}})$ satisfying $g(2-s)=g^T(s)^{-1}$ such that

\bea
\a'= g\a g^{-1}-\frac{1}{2}\frac{dg}{ds}g^{-1} \\
\b'= g\b g^{-1} \\
v'=g(0)v
\eea

The main result of reference \cite{don} is that given a Nahm complex $(\a,\b,v)$ there exist a solution to Nahm's equation $T_i(s)$ and a weight vector $v'$ such that $T_1v'=\frac{(k-1)}{4}v'$ for $T_1$ the residue at $s=0$ of $T_1(s)$ such that $(T_i,v')=(\a',\b',v')$ for some $g\in GL(k,{\mathbb{C}})$.

In particular, notice that $(T_i,v_1)$ and $(T_i,v_2)$ with $T_i$ solutions to Nahm's equations (Th.\ref{h}.\ref{h1}) and $v_2=e^{i\phi}v_1$, define two Nahm's complex $(\a,\b,v_1)$, $(\a,\b,v_2)$ which are not gauge equivalent. In fact, if $g(0)=e^{i\phi}$ and we require $g(2-s)=g^T(s)^{-1}$ we get $g(s)=e^{i\phi(s-1)}$ that induce the extra piece in $\a'$, namely

\be
\a'=\a-\frac{i}{2}\phi
\ee

\noindent this explains the $S^1$ bundle nature of ${\cal{M}}_k$.

With these pieces we can define the rational function $S_m(z)$. Given a Nahm complex $(\a,\b,v)$ we define the symmetric matrix $B$ as $\b(1)$ and a vector $w\in{\mathbb{C}}^k$ as $u(1)$ for $u(s)$ solution to

\be
\left(\frac{1}{2}\frac{d}{ds}+\a\right)u=0
\ee

\noindent which is the equation (\ref{hne1}) for a $\mathbb{C}^k$ valued function, and such that $u(s)s^{-(k-1)/2}\rightarrow v$ for $s\rightarrow 0$. The rational function $S_m(z)$ is then defined by

\be
S_m(z)=w^T\frac{1}{B-z{\mathbb{I}}}w
\ee

It is interesting to see how the scattering function is transformed by the action of the group $SO(2)\times {\bf R}\times U(1) \times {\bf C}$, where $\bf R$ denotes a translation in the $x_1$ direction, $\bf C$ gives a translation in the orthogonal plane, $U(1)$ represents the phase change and $SO(2)$ represents the subgroup of $SO(3)$ fixing the preferred direction in $\mathbb{R}^3$. Let

\be
(\l,\m,\n)\in SO(2)\times {\bf C}^*\times {\bf C}\cong SO(2)\times {\bf R}\times U(1) \times {\bf C}
\ee

\noindent where $\l$, $\m$ and $\n$ are complex numbers, $|\l|=1$, $\m\neq 0$ and $\log|\m|\in{\bf R}$, $\m/|\m|\in U(1)$. Then the scattering function transforms as

\be
\label{tms}
\m^{-2}\l^{-2k}S\left(\l^{-1}(z-\n)\right)
\ee

It is posible to prove \cite{don} that $S_m(z)$ depends only on the equivalence class of Nahm complex $(\a,\b,v)$. For instance, in the simplest case $k=1$ we can take $T_i(s)$ as arbitrary constants. the $B=b$ for $b$ an arbitrary complex number and $u(s)=ae^{i\a s}$ where $a$ and $\a$ are also arbitrary complex numbers. Depending on $v$ the phase of $a$ will be fixed. The function is simply

\be
\label{s1}
S_1(z)=\frac{a}{z-b}
\ee

A change in $S^1:v\rightarrow e^{i\phi}v$ will induce a change of $a$ to $e^{i\phi}a$ in (\ref{s1}). The moduli space ${\cal{M}}_1={\mathbb{R}}^3\times S^1$ is parametrized by $(|a|,b,arg(a))$ where again we have used the isomorphism ${\mathbb{R}}^3\sim {\mathbb{C}}\times{\mathbb{R}}$.

Next let us study the change of the scattering function for the non-commutative case.

\subsection{The non-commutative case}

In order to fix ideas, let us consider a constant $B$ field in the $(1,2)$ directions. The solution to the deformed equations (\ref{dene}) is then

\be
T_i^{NC}=T_i^C+\theta s\d_{i3}
\ee

Next we need to choose the isomorphism $\mathbb{R}^3\sim\mathbb{C}\times\mathbb{R}$. We can consider various possibilities, namely 

\bea
\label{i1}(x_1,x_2,x_3)\rightarrow(x_1,x_2+ix_3)\\
\label{i2}(x_1,x_2,x_3)\rightarrow(x_2,x_1+ix_3)\\
\label{i3}(x_1,x_2,x_3)\rightarrow(x_3,x_1+ix_2)
\eea

\noindent depending on what direction in $\mathbb{R}^3$ we map into $\mathbb{R}$ in $\mathbb{C}\times\mathbb{R}$, which can be also seen as the plane where the $B$ field lies once the complex structure has been fixed. This choice will produce different deformations in equations (\ref{hne1}) and (\ref{hne2}). Choosing the isomorphism (\ref{i3}) we get

\bea
\label{hnei31}\frac{d\b}{ds}+2[\a,\b]=0\\
\label{hnei32}\frac{d}{ds}\left(\a+\a^*\right)+2\left([\a,\a^*]+[\b,\b^*]\right)+\theta=0
\eea

\noindent while choosing the isomorphism (\ref{i1}) will produce the deformation

\bea
\label{hnei11}\frac{d\b}{ds}+2[\a,\b]+i\theta=0 \\ 
\label{hnei12}\frac{d}{ds}\left(\a+\a^*\right)+2\left([\a,\a^*]+[\b,\b^*]\right)=0
\eea

Let us consider first the deformation (\ref{hnei31}), (\ref{hnei32}). This corresponds to

\bea
\a^{NC}=\a^C+\frac{i\theta s}{2},\qquad \b^{NC}=\b^C
\eea

If we now define the scattering function in the same way as for the commutative case, we get

\bea
B^{NC}=B^C,\qquad w^{NC}=w^Ce^{-i\theta/2}
\eea

\noindent which, in turn, implies

\be
\label{snc1}
S_m^{NC}=S_m^C(z)e^{-i\theta}
\ee

For the simplest case $k=1$ we get

\be
\label{snc2}
S_1^{NC}(z)=\frac{ae^{-i\theta}}{z-b}
\ee

This in particular means that the moduli spaces ${\cal{M}}_k^{NC}$ and ${\cal{M}}_k^C$ are isomorphic. For $k=1$ the ismorphism maps the point $(|a|,b,arg(a))\in\mathbb{R}^3\times S^1$ into $(|a|,b,arg(a-i\theta))$.

If we had chosen the deformation of (\ref{hnei11}), (\ref{hnei12}) the scattering function obtained would have been

\be
S_1^{NC}(z)=\frac{a}{z-b-\frac{1}{2}\theta}
\ee 

\noindent i.e. a different isomorphism between ${\cal{M}}_k^{NC}$ and ${\cal{M}}_k^C$. However, both isomorphisms can be easily related by the action of the group of symmetries acting on $S_k(z)$, given by equation (\ref{tms}).

There is, however, an interesting thing we can derive form (\ref{snc1}) and (\ref{snc2}). Namely, changes of the non-commutative parameter $\theta$ define motions in the $S^1$ fiber of ${\cal{M}}_k^{NC}$. This means that the generator of changes of $\theta$ is precisely the electric charge of the monopole. Formally we can write

\be
\label{qms}
e^{i\theta Q_e}|m\rangle_{B=0}=|m\rangle_{B=\theta}
\ee

\noindent where $Q_e$ is the electric charge and $|m\rangle_{B=\theta}$ is the quantum state of the monopole in presence of the $B$ field. Since $\pi_1({\cal{M}}_k)=\mathbb{Z}$ we can consider the change of the sate $|m\rangle_B$ going from $B=0$ to $B=\theta=2\pi$ corresponding to a non contractible loop on ${\cal{M}}_k$. In the instanton $\Theta$ vacua we expect to get

\be
e^{2\pi iQ_e}|m\rangle=e^{-ik\Theta}|m\rangle
\ee

\noindent for the monopole of charge equal to $k$. Replacing in (\ref{qms}) we get

\be
e^{i\theta\frac{\Theta k}{2\pi}}|m\rangle_{B=0}=|m\rangle_{B=\theta}
\ee

\noindent i.e. 

\be
i\frac{\pd}{\pd\theta}|m\rangle_{B=\theta}=k\frac{\Theta}{2\pi}|m\rangle_{B=\theta}
\ee

Let us finish this section with the following general remarks.

An interesting difference between the BPS monopole and instanton moduli spaces is that in the instanton case the moduli space is singular, corresponding to small instanton configurations of size equal to zero. A resolution ot this singularity was suggested by Nakajima in reference \cite{naka} and corresponds to the replacement of the real moment map $\m_R=0$ in the ADHM construction by $\m_R=constant$. Nekrasov and Schwarz noticed in \cite{ns} that this resolution of the small instanton singularity of the instanton moduli space exactly corresponds to the non-commutative deformation of the ADHM real moment map being the constant deformation the non-commutative parameter.

In the monopole case the situation is different. The moduli space ${\cal{R}}_k$ of rational functions of degree $k$ is already a Hilbert scheme $X^{[k]}$, with $X$ the moduli space $\mathbb{R}^3\times S^1$ of a $k=1$ monopole. The Hilbert scheme $X^{[k]}$ is a desingularization of ${\cal{S}}^k(X)$ that would be the natural moduli space os $k$-monopoles interpreted as a set of $k$ different particles.

For the case $k=2$, $X^{[2]}$ can be visualized as the blow up of $X\times X$ along the diagonal $\Delta$, divided by $\mathbb{Z}_2$. Thus, the Hilbert scheme blows up the singularities corresponding to two colliding points. The way this is done is taking into account the way the two colliding points approach each other.

In the non-commutative case, the real moment map $\m_R$, as defined by equation (\ref{hne2}), is modified to $\m_R=-\theta$, as given in (\ref{hnei32}). However, what is interesting is that in the monopole case we do not need this deformation of $\m_R$ to desingularize the moduli space, which is already the smooth Hilbert scheme $\left[\mathbb{R}^3\times S^1\right]^{[k]}$. The physical reason for this is that even without $B$ field, the elementary constituents of a $k$-monopole are at short distances delocalized; they spread out, in the simplest $k=2$ case, in the divisor blowing up $\Delta$.

A nice string version of Nahm's construction is due to Diaconescu \cite{dia}. In this approach one considers two parallel D3-branes in type IIB and a set of $k$ D1-branes in between. The $(k\times k)$ Nahm matrices $T_i(s)$ are now interpreted as the transversal position of the D1-branes in the D3 world volume. Nahm's equations for the $T_i(s)$ become the BPS vacuum conditions for the D1-branes. By applying T-duality in these transversal directions we get a set of D4-branes from the D-strings, and we transform the D3-branes into D0-branes. The coordinates $T_i$ become now the gauge field on the D4-brane, that we can simply identify with Donladson's connection (\ref{dc}). 

Since we are working with D0 and D4-branes, we are in type IIA and it makes sense to think about a M-theory lift. In fact, the two D0-branes can have charge that we can interpret as momentum in the extra eleven dimension. To the center of mass of these two D0-branes we can associate a vector $v$ with $|v|=1$ defining the ``position'' in the extra $S^1$. It is natural to identify this vector $v$ with the one used in Donaldson's definition of Nahm's complex\footnote{More precisely, with the $S^1$ phase of the ${\mathbb{C}}^k$ vector $v$ of Nahm's complex}.

\section*{Acknowledgements}

One of us, C.G., thanks Nankai Institute of Mathematical Physics and the E.~Schr\"oedinger Institute for hospitality during the completion of this work. This research was possible thanks to the grant AEN2000-1584.



\begin{thebibliography}{99}

\bibitem{mm} R.~Minasian and G.~Moore,
                        {\em K-theory and Ramond-Ramond Charge},
                        J.\ High Energy Phys.\ {\bf 11} (1997) 002, 
                        {\tt hep-th/9710230}.


\bibitem{witten2} E.~Witten,
                  {\em D-branes and K-theory},
                  J.\ High Energy Phys.\ {\bf 12} (1998) 019,
                  {\tt hep-th/9810188}.

\bibitem{mw} G.~Moore and E.~Witten,
                        {\em Self-duality, Ramond-Ramond Fields, and K-theory},
                        J.\ High Energy Phys.\ {\bf 05} (2000) 032,
                        {\tt hep-th/9912279}.


\bibitem{hora} P.~Ho\v rava,
               {\em Type-IIA D-branes, K-theory and Matrix Theory},
               Adv.\ Theor.\ Math.\ Phys.\ {\bf 2} (1998) 1373,
               {\tt hep-th/9812135}.


\bibitem{sen2} A.~Sen,
               {\em Tachyon condensation on the brane anti-brane system},
               J.\ High Energy Phys.\  {\bf 08} (1998) 012,
               {\tt  hep-th/9805170}\\
               {\em Universality of the tachyon potential},
                J.\ High Energy Phys.\ {\bf 12} (1999) 027,
               {\tt  hep-th/9911116}.


\bibitem{ces} C.~G\'omez,
                        {\em A Comment on the K-theory Meaning of the Topology Gauge Fixing},
                        {\tt hep-th/0104211}.


\bibitem{sen} A.Sen,
              {\em Kaluza-Klein Dyons in String Theory},
              Phys.\ Rev.\ Lett.\ {\bf 79} (1997) 1619-1621, 
              {\tt hep-th/9705212}.


\bibitem{ghm} R.~Gregory, J.A.~Harvey and G.~Moore,
              {\em Unwinding Strings and T-duality of Kaluza-Klein and H-Monopoles},
              Adv.\ Theor.\ Math.\ Phys.\ {\bf 1} (1997) 283-297,
              {\tt hep-th/9708086}.


\bibitem{dmw1} E.~Diaconescu, G.~Moore and E.~Witten,
               {\em $E_8$ Gauge Theory, and a Derivation of K-theory from M-theory},
               {\tt hep-th/0005090}. 


\bibitem{gr} G.W.~Gibbons and P.J.~Ruback,
                    {\em Winding Strings, Kaluza-Klein Monopoles and Runge-Lenz Vectors},
                    Phys.\ Lett.\ {\bf B215} (1988) 653-656.


\bibitem{mms} J.~Maldacena, G.~Moore and N.~Seiberg,
                          {\em D-brane Charges in Five Brane Backgrounds},
                          J.\ High Energy Phys.\ {\bf 10} (2001) 005,
                          {\tt hep-th/ 0108152}



\bibitem{sor} R.~Sorkin,
              {\em Kaluza-Klein Monopole},
              Phys.\ Rev.\ Lett.\ {\bf 51} 2 (1983) 87



\bibitem{gp} D.J.~Gross and M.J.~Perry,
             {\em Magnetic Monopoles in Kaluza-Klein Theories},
             Nucl. Phys. {\bf B226} (1983) 29.


\bibitem{gh} G.~Gibbons and S.~Hawking,
             {\em Classification of gravitational instanton symmetries},
             Comm.\ Math.\ Phys.\  {\bf 66} (1979) 291-310.


\bibitem{witten1} E.~Witten,
                               {\em Dyons of charge $\frac{e\theta}{2\pi}$. },
                               Phys.\ Lett.\ {\bf B86} (1979) 283-287.



\bibitem{gs} M.~Green and J.H.~Schwarz,
             {\em Anomaly Cancellation in Supersymmetric $D=10$ Gauge Theory and Superstring Theory},
             Phys.\ Lett.\ {\bf B149} (1984) 117-122.


\bibitem{town} P.~Townsend,
                        {\em The eleven dimensional membrane revisited},
                        Phys.\ Lett.\ {\bf B350} (1995) 184-187,
                       {\tt  hep-th/9501068}.\\
                        C.~Hull,
                        {\em Gravitational duality, branes and charges},
                        Nucl.\ Phys.\ {\bf B509} (1998) 216-251,
                        {\tt  hep-th/9705162}.

\bibitem{dl} M.J.~Duff and J.X.~Lu,
             {\em Black and Super p-branes in diverse dimensions},
             Nucl.\ Phys.\ {\bf B416} (1997) 301-334,
             {\tt hep-th/9306052}.
                        

\bibitem{os} K.~Olsen and R.J.~Szabo,
                      {\em Constructing D-branes from K-theory},
                      {\tt hep-th/9907140}.


\bibitem{yi} Yosuke Imamura,
                         {\em Born-Infeld action and Chern-Simons term from Kaluza-Klein monopole in M-theory,}
                         Phys.\ Lett.\ {\bf B414} (1997) 242-150,
                         {\tt hep-th/9706144}


\bibitem{hw} P.~Ho\v rava and E.~Witten,
             {\em Heterotic And Type I String Dynamics From Eleven Dimensions},
             Nucl.\ Phys.\ {\bf B460} (1996) 506,
             {\tt hep-th/9510209};\\
             {\em Eleven-Dimensional Supergravity On A Manifold With Boundary},
             Nucl.\ Phys.\ {\bf B475} (1996) 94-114,
             {\tt hep-th/9603142}.


\bibitem{witten3} E.~Witten, 
                  {\em On Flux Quatization in M-theory and the effective action},
                  J.\ Geom.\ Phys.\ {\bf 22} (1997) 1-13,
                  {\tt hep-th/9609122}.


\bibitem{dmw2} E.~Diaconescu, G.~Moore and E.~Witten,
               {\em A Derivation of K-theory from M-theory},
               {\tt hep-th/0005091}.


\bibitem{ah} M.F.~Atiyah and N.J.~Hitchin,
                      {\em The geometry and dynamics of magnetic monopoles},
                      {\tt Princeton Univ.\ Press}, Princeton, NJ, 1988.


\bibitem{nahm} W.~Nahm,
                            {\em The construction of all self-dual monopoles by the ADHM method},
                            in Monopoles in Quantum Field Theory, Proceedings of the monopole meeting in Trieste 1981, 
                           World Scientific, Singapore (1982);\\
                            {\em All self-dual multimonopoles for arbitrary gauge group},
                            (Preprint), TH 3172-CERN (1981);\\
                            {\em The algebraic geometry of multimonopoles},
                            (Preprint), Physikalisches Institut, University of Bonn.



\bibitem{adhm} M.F.~Atiyah, V.G.~Drinfield, N.J.~Hitchin and Yu I.~Manin,
                           {\em Construction of instantons},
                           Phys.\ Lett.\ {\bf A65} (1978) 185-187. 




\bibitem{hitchin} N.~Hitchin,
                  {\em On the construction of monopoles},
                  Comm.\ Math.\ Phys.\ {\bf 89} (1983) 145-190.




\bibitem{bak} D.~Bak,
                        {\em Deformed Nahm equation and a noncommutative BPS monopole},
                        Phys.\ Lett.\ {\bf B471} (1999) 149,
                        {\tt hep-th/9910135}.



\bibitem{gn} D.~Gross and N.~Nekrasov,
                      {\em Monopoles and strings in non-commutative gauge theory},
                      J.\ High Energy Phys.\ {\bf 07} (2000) 034,
                     {\tt hep-th/0005204}




\bibitem{don} S.K.~Donaldson,
                        {\em Nahm's equations and the classification of monopoles},
                        Comm.\ Math.\ Phys.\ {\bf 90} (1984) 387-407;



\bibitem{kron} N.J.~Hitchin, A.~Karlhede, U.~Lindstr\"om and M.~Ro\v cek,
                         {\em Hyperk\"ahler metrics and supersymmetry},
                          Comm.\ Math.\ Phys.\ {\bf 108} (1987) 535-589;\\
                          P.B.~Kronheimer,
                         {\em The construction of ALE spaces as hyper-k\"ahler quotients},
                         J.\ Diff.\ Geom. {\bf 29} (1989) 665-683.



\bibitem{naka} H.~Nakajima,
                          {\em Heisenberg algebra and Hilbert schemes of points on porjective surfaces},
                          {\tt alg-geom/9507012};\\
                          {\em Lectures on Hilbert schemes of points on surfaces}, 
                           www.kusm.kyoto-u.ac.jp/$\sim$nakajima


\bibitem{ns} N.~Nekrasov and A.~Schwarz,
                      {\em Instantons on noncommutative $\mathbb{R}^4$ and $(2,0)$ superconformal six dimensional theory},
                      {\tt hep-th/9802068}.


\bibitem{dia} D.~Diaconescu,
                        {\em D-branes, monopoles and Nahm equations},
                        Nucl.\ Phys.\ {\bf B503} (1997) 220-238,
                        {\tt hep-th/9608163}.


\end{thebibliography}
\end{document}